\documentclass[acmsmall,screen]{acmart}
\usepackage{amsmath,amsfonts}
\usepackage{algorithmic}
\usepackage{algorithm}
\usepackage{array}
\usepackage{textcomp}
\usepackage{stfloats}
\usepackage{url}
\usepackage{verbatim}
\usepackage{graphicx}
\usepackage{soul}

\usepackage{booktabs}
\usepackage{subfigure}

\usepackage{xcolor}
\usepackage{colortbl}

\usepackage{multirow}
\usepackage{makecell}
\usepackage{diagbox}
\usepackage{enumitem}
\usepackage{blindtext}
\usepackage[english]{babel}
\usepackage{csquotes}
\usepackage{gensymb} 
\usepackage{marvosym}
\usepackage{pifont}
\usepackage{threeparttable}
\usepackage{hyperref}
\renewcommand\footnotetextcopyrightpermission[1]{}
\AtBeginDocument{%
  }

\author{Dingcui Yu}
\authornote{Dingcui Yu, Yunpeng Song and Yiyang Huang have equal contributions to this work.}
\email{dingcuiy@gmail.com}
\orcid{0009-0006-5344-2031}
\affiliation{%
  \institution{College of Computer Science and Technology, East China Normal University}
  \city{Shanghai}
  \country{China}
}

\author{Yunpeng Song*}
\email{51205901045@stu.ecnu.edu.cn}
\orcid{0009-0006-3822-3551}
\affiliation{%
  \institution{College of Computer Science and Technology, East China Normal University}
  \city{Shanghai}
  \country{China}
  }

\author{Yiyang Huang*}
\email{51215901085@stu.ecnu.edu.cn}
\affiliation{%
  \institution{College of Computer Science and Technology, East China Normal University}
  \city{Shanghai}
  \country{China}
  }

\author{Yumiao Zhao}
\email{zhaoyumiao99@gmail.com}
\orcid{0000-0001-7803-2887}
\affiliation{%
  \institution{College of Computer Science and Technology, East China Normal University}
  \city{Shanghai}
  \country{China}
  }
  
\author{Yina Lv}
\email{elainelv95@gmail.com}
\orcid{0000-0003-3971-3123}
\affiliation{%
  \institution{School of Informatics, Xiamen University}
  \city{Xiamen}
  \country{China}
  }

\author{Chundong Wang}
\email{toast-lab@outlook.com}
\orcid{0000-0002-5570-0018}
\affiliation{%
  \institution{ShanghaiTech University}
  \city{Shanghai}
  \country{China}
  }

  \author{Youtao Zhang}
\email{youtao@pitt.edu}
\orcid{0000-0001-8425-8743}
\affiliation{%
  \institution{Computer Science Department, University of Pittsburgh}
  \city{Pittsburgh}
  \country{USA}
  }

\author{Liang Shi}
\authornote{Liang Shi and Yina Lv are the corresponding authors.}
\email{shi.liang.hk@gmail.com}
\orcid{0000-0002-9977-529X}
\affiliation{%
  \institution{College of Computer Science and Technology, East China Normal University}
  \city{Shanghai}
  \country{China}
  }

\ccsdesc[500]{Hardware~Power and energy}
\ccsdesc[300]{Software and its engineering}
\ccsdesc[500]{Information systems~Storage power management}
\ccsdesc[300]{Information systems~Storage architectures}
\ccsdesc[500]{Information systems~Flash memory}

\renewcommand{\shortauthors}{Dingcui et al.}

\begin{document}

\title{Waltz: Temperature-Aware Cooperative Compression for High-Performance Compression-Based CSDs}


\begin{abstract}
Data compression is widely adopted for modern solid-state drives (SSDs) to mitigate both storage capacity and SSD lifetime issues.
Researchers have proposed compression schemes at different system layers, including device-side solutions like CCSDs (\underline{c}ompression-based \underline{c}omputational S\underline{SD}s) and compression supported by host-side, like F2FS (flash-friendly file system).
We conduct quantitative studies to understand how host-side and device-side compression schemes affect the temperature and performance of SSD-based storage systems.
From our experiments, device-side compression, facilitated by a hardware compression engine, can raise the temperature of CCSDs to intolerable levels, resulting in throttling and service shutdown.
In contrast, host-side compression causes software-stack overhead, which often results in large performance degradation and resource consumption.
To ensure efficient data compression with high performance and better temperature control, we propose Waltz, a temperature-aware cooperative compression method that schedules (de)compression tasks at the host and device sides by monitoring device temperature.
Furthermore, we introduce two variants (Waltzs and Waltzp) for space and performance optimization, respectively.
Waltz is implemented within F2FS, achieving high performance while extending SSD lifetime and preventing overheating-induced in-flight shutdowns.

\end{abstract}

\maketitle
\begin{sloppypar}
\section{Introduction}
Modern applications, such as large language models \cite{alizadeh2024llmflashefficientlarge} and streaming services \cite{8954623}, increasingly demand large data storage.
NAND flash-based SSDs (solid-state drives), while being widely adopted in modern computer systems, have a limited endurance due to the out-of-place update characteristic. 
Data compression is a simple yet effective solution that has been adopted for SSDs \cite{tate2018ibm,264858,zuck2014compression}. 
By reducing the amount of data to be written to SSDs, data compression effectively mitigates write amplification and prolongs SSD lifetime \cite{zuck2014compression}.
Data compression saves storage space and thus reduces storage costs \cite{tate2018ibm}. 
Alternatively, the saved space can be used to prolong the lifetime of SSDs or improve performance.

Existing compression methods are deployed either on the device side \cite{tate2018ibm,264858,zuck2014compression,song2023f2fs} or on the host side \cite{cF2FS,e2compr,gao2019erofs,hu2019qzfs}.
The device-side compression methods are often combined in CSDs (computational SSDs) \cite{li2021glist},
referred to as compression-based computational SSDs (CCSDs)~\cite{CSD3000}.
Some mainstream companies, such as ScaleFlux, Samsung, and Dell, have released their computational storage products \cite{CSD3000,dell,samsung} that integrate 
ASIC or FPGA processing engines inside the storage controller.
The host-side compression methods utilize host CPUs to perform (de)compression and are often integrated within the file system.
For example, F2FS (flash-friendly file system) \cite{lee2015F2FS} has supported compression in the Linux kernel since version 5.6.

However, the above single-sided compression methods cannot achieve optimal performance with flexible temperature management.
Adopting data compression in file systems introduces non-negligible software-stack overhead.
We observed a throughput degradation, especially for random writes, which experiences a 50$\times$ performance drop (see \S~\ref{moti:performance}). 
On the other hand, the device-side compression mitigates performance drop through hardware compression engines.
However, it faces \ul{\bf overheating problem} when processing compressible data continuously. 
A CCSD may be overheated with ordinary write operations in about eight minutes (see \S~\ref{sec:te}).
Due to small form factor, cost, and noise considerations, current CSD products lack in-package cooling and often {\em throttle} compression, i.e., reduce (de)compression throughput when overheating occurs.
In this case, the device-side compression is temporarily closed.
A system restart is required to re-enable the device-side compression.

The above discussion motivates the design of a data compression scheme that combines both device-side and host-side compression. 
Typically,
device-side and host-side compression schemes usually operate at different granularity.
Host-side compression typically employs large granularity (e.g., minimum 16KB) to obtain higher compression ratios.
In contrast, device-side compression processes fixed-size blocks (e.g., 4KB) independently, then concatenates multiple compressed blocks to fit physical block sizes.
In addition, the compression algorithms used by the device-side and host-side compression schemes are different.
Therefore, a simple integration faces challenges, including misaligned compression granularity, incompatible algorithms, and complicated metadata management.

In this paper, we propose Waltz\footnote{Waltz is a graceful and romantic duo ballroom dance presented by two dancers in cooperation.}, a temperature-aware cooperative compression method to achieve better temperature control and performance improvements.
Specifically, Waltz schedules (de)compression tasks between the host and the device based on system performance demands and the temperature status of the CCSD. 
It consists of three building blocks: cooperative compression framework (CCF), temperature-aware compression scheduling (TCS), and Waltz configuration selection (TWINS). 
The CCF enables collaborative (de)compression between F2FS and CCSD. 
The TCS adapts task scheduling to the CCSD’s temperature.
The TWINS allows users to prioritize optimization goals, choosing between reduced write amplification factor (WAF) or improved performance and introducing the two variants (Waltzs and Waltzp) respectively.
Furthermore, we conducted a case study and proposed on-demand space allocation (OSA), a method that uses the space saved by compression as the reserved space to further improve system performance and flash lifetime.

We evaluate Waltz’s effectiveness in controlling CCSD temperature and its performance against existing methods.
Using a testbed with a CCSD emulator built from a real-world temperature model, we tested Waltz with both synthetic and real-world workloads.
Our experiments show that while the vanilla CCSD (i.e. the baseline) suffers from overheating and system shutdowns, Waltz effectively manages CCSD temperature, preventing thermal emergencies under all workloads. 
This is achieved by cooperatively offloading (de)compression tasks to the host.
Beyond temperature control, Waltz significantly boosts system throughput. 
The cooperative compression framework itself, represented by the TCS scheme, delivers substantial write and read throughput improvements of up to 391.5\% and 627\% over the baseline, respectively, by eliminating the need for throttling. 
The OSA scheme, a core component of Waltz, further enhances performance by effectively using saved space to optimize the segment cleaning in F2FS. 
This is proven by Waltzs and Waltzp achieving up to 95.5\% and 23.6\% higher write throughput than the TCS scheme, respectively, when segment cleaning operations are activated. 
Overall, our results demonstrate that Waltz is a robust solution that simultaneously solves the overheating problem and achieves superior performance.

The main contributions are as follows.
\begin{itemize}
    \item We propose to cooperatively schedule (de)compression requests from different I/O applications at either the host side or the device side. This is achieved by detecting the temperature impacts of (de)compression operations, and then making I/O scheduling according to the requirements of different applications as well as the temperature status of the CCSDs.
    \item We propose cross-layer scheduling to optimize the tradeoffs among high performance, low WAF {\footnote{WAF indicates the write amplification factor, which is defined as \\ $\frac{\textnormal{Physical writes to NAND flash (after compression)}}{\textnormal{Host logical writes (before compression)}}$}}, and temperature control. 
    In particular, we exploit the saved space from compression to enlarge the reserved space (RS) of F2FS for optimized performance and WAF.
    \item We implement Waltz with F2FS. Experimental results show that Waltz optimizes performance and lifetime to build a cost-efficient storage system without in-flight shutdown.
\end{itemize}

The rest of the paper is organized as follows. 
\S~\ref{sec:back} introduces the background.
\S~\ref{sec:obs} characterizes the system impacts of data compression on the host and device sides and poses the problem of current design.
\S~\ref{sec:coco} presents the design and implementation of Waltz.
\S~\ref{sec:case} presents the case study that shows the possible usage of space saved by compression.
\S~\ref{sec:eva} discusses the the evaluation results of Waltz.
Finally, \S~\ref{sec:related} and \S~\ref{sec:conclusion} presents the related works and gives the conclusion respectively.

\section{Background}\label{sec:back}
\subsection{Flash-based SSDs and Data Compression}
NAND flash-based SSDs are widely adopted in modern computer systems to meet the increasing demand for large-capacity data storage \cite{hu2019qzfs,alizadeh2024llmflashefficientlarge}.
An SSD module typically consists of two main components: a NAND flash chip array for data storage, and a controller for data management \cite{song2023decc}.
Inside the chip, there are hundreds or thousands of blocks.
Each block consists of hundreds or thousands of pages.
One flash page is typically configured from 4KB to 16KB.
Due to the out-of-place update property of NAND flash memory, flash pages cannot be rewritten until they are erased via a garbage collection (GC) that reclaims space by erasing blocks.
For each GC, one block will experience one program and erase (P/E) cycle.
The SSD often maintains an over-provision (OP) space that is invisible to users and used to store valid data migrated during GC.
However, due to the manufacturing process, the endurance of flash blocks is limited by the number of P/E cycles, which constrains the total written volume for an SSD \cite{song2023adaptive}.
To optimize the lifetime and I/O performance of SSDs, compression techniques are applied to the SSD by reducing the amount of data written to flash memory via compression algorithms \cite{cF2FS}. 
For example, when 16KB of data is compressed into 8KB (i.e., the compression ratio is 2), the written data volume is reduced by half.
Furthermore, the system layer adapts to SSD properties by implementing techniques tailored to flash memory, such as deploying a log-structured file system (LFS) like F2FS.

\subsection{Device-Side Compression}
Early device-side compression schemes implement software compression in hardware \cite{zuck2014compression}, which offloads the software stack overhead through hardware acceleration.
Recent studies focus on hardware-implemented (ASIC or FPGA) compression engines in CCSDs \cite{zhanghotstorage20Rethink}.
Hardware-implemented compression engines can reduce the compression latency to about 10 microseconds with negligible impact on performance \cite{CSD3000}.

\begin{figure}[htbp]
\centerline{\includegraphics[width=0.7\linewidth]{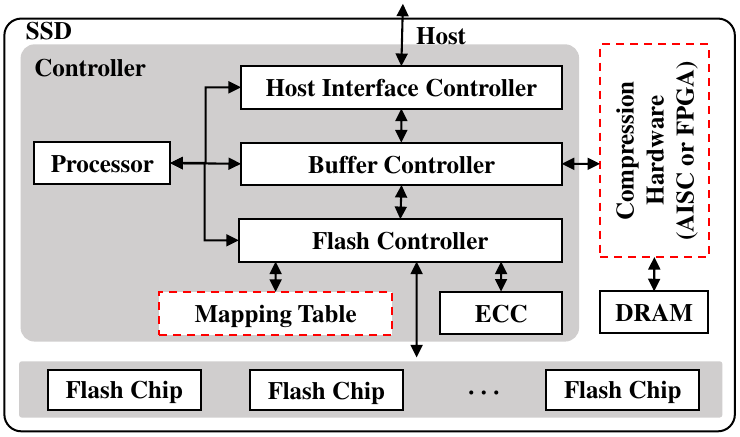}}
\caption{The architecture of CCSD.}
\label{fig:overview}
\end{figure}

Fig. \ref{fig:overview} shows the architecture of CCSD.
Compared with traditional SSDs, the added and modified modules are highlighted with red dashed boxes.
The hardware-implemented (ASIC or FPGA) scheme compresses inputs with a fixed length, e.g., 4KB. 
Suppose that the user needs to write 16KB data to the storage. 
The CCSD receives and splits the data into four 4KB chunks and compresses each 4KB individually. 
In addition, the hardware-implemented compression produces variable-length output data. 
Consequently, compressed data accumulate in the DRAM of the SSD controller and are flushed to the flash memory in physical block (page) size \cite{zuck2014compression}. 
The logical to physical address mapping table has been modified to accommodate variable-length data \cite{zuck2014compression}.

Data compression exposes a new design opportunity, i.e., how to utilize the space saved by compression.
With data compression, the size of the logical storage address space exceeds that of the physical space. 
The CCSD decouples the logical space from the physical space such that an upper-layer software can observe a larger logical space \cite{zhanghotstorage20Rethink}.
By providing an interface to detect the actual physical space usage, the CCSD allows the software to exploit the saved space. 
In this work, we discuss potential applications of compressed space and present a case study to illustrate the benefits of space compression.

Due to small form factors, cost, and noise considerations, current CSD products lack in-package cooling. 
However, it often integrates a temperature sensor whose readings can be sent to the SSD controller.
The CCSD uses a simple throttling strategy to address the \textbf{\underline{overheating problem}}, however, this will significantly reduce (de)compression throughput \cite{CSD3000,xu2019lessons} (see \S~\ref{sec:te}). 

It is challenging to reduce the operating temperature of CSD products without in-package cooling. 
Our studies showed that CSDs still face fast temperature increases even when they are deployed in rooms with aggressive cooling, i.e., the heat generated from computation outpaces the heat dissipated to cool ambient air. 
The cluster organization of SSDs in data centers may worsen the situation, e.g., Xu et al. \cite{xu2019lessons} showed that the temperature of an idle SSD can be heated up by neighboring SSDs in such a setting.



\subsection{Host-Side Compression}\label{sec:hsc}
Multiple software-implemented compression methods have been developed on the host side \cite{cF2FS,gao2019erofs,sears2008rose}.
F2FS is an LFS (Log-Structured File System) that supports compression since Kernel version 5.6.0 \cite{f2fsgit}.
In the following discussion, we use F2FS as the baseline and extend it to support our proposed cooperation compression. 
Our design can be extended to other systems as well. 

\begin{figure}[htbp]
\centerline{\includegraphics[width=0.7\linewidth]{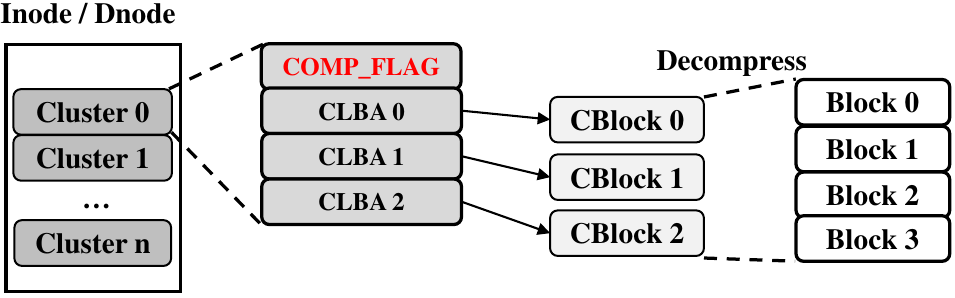}}
\caption{The cluster layout of the compressed data in F2FS.}
\label{fig:cluster}
\end{figure}

F2FS supports data compression by compressing data at \ul{cluster} (at least 16 KiB) granularity.  
Fig. \ref{fig:cluster} illustrates the index structure of the compressed blocks. 
From the figure, a cluster contains several contiguous blocks (with the number being configurable).
F2FS uses the address of the first block (COMP\_FLAG in the figure) to indicate if the cluster is compressed, and the other addresses (CBlock in the figure) to index the compressed blocks. 
This compressed data layout leads to the loss of the boundaries of the original uncompressed blocks.
When servicing user I/O requests, F2FS reads and decompresses the compressed cluster to restore the boundaries across the blocks, and then returns the requested data block based on the mapping.

To reclaim the space occupied by invalid data from out-of-place updates, F2FS adopts a segment cleaning (SC) scheme.
While the SC process for F2FS is the same as the GC process for SSD, the unit of SC is a segment.
To mitigate the overhead, F2FS allocates reserved space (RS) to assist the SC process \cite{lee2015F2FS}.
The RS space is used to store valid data for the victim segment during SC.
To alleviate the performance impact of SC, F2FS supports in-place updates called thread logging \cite{lee2015F2FS}.
F2FS activates the SC process based on the amount of free space left.
When the size of the free space is less than a threshold, thread logging is triggered, and vice versa.

\section{Problem Statement and Empirical Study} \label{sec:obs}
To characterize the system impacts of data compression on the host and device sides, we conduct experiments on real CCSDs with F2FS and evaluate the \textit{temperature change in CCSD}, the \textit{performance}, and the \textit{CPU and memory overhead of F2FS with compression}.
We use FIO \cite{fio} in this empirical study. 
More details of the setting can be found in~\S~\ref{sec:eva}.

\subsection{The Temperature Variation in CCSD}\label{sec:te}
The hardware-implemented compression engine in CCSDs, while achieving high performance in data compression, faces the temperature problem. 
The default strategy for overheating is to throttle throughput, which degrades performance or even shuts down the system~\cite{xu2019lessons}.
To quantify the overheating problem, we evaluate the temperature change of an ASIC-based CCSD with or without compression. 
All tests are performed at the room temperature of 23$\degree$C. 
We start the experiment with an idle CCSD at room temperature, and then issue sequential writes of 128KB data blocks for 20 minutes.
We set the number of threads to 8 to maximize the CCSD throughput, and the compression ratio to 1 to avoid performance variation due to compression.
We collect the temperature every two seconds by reading the integrated temperature sensor in the CCSD.

\begin{figure}[htbp]
\centerline{\includegraphics[width=0.7\linewidth]{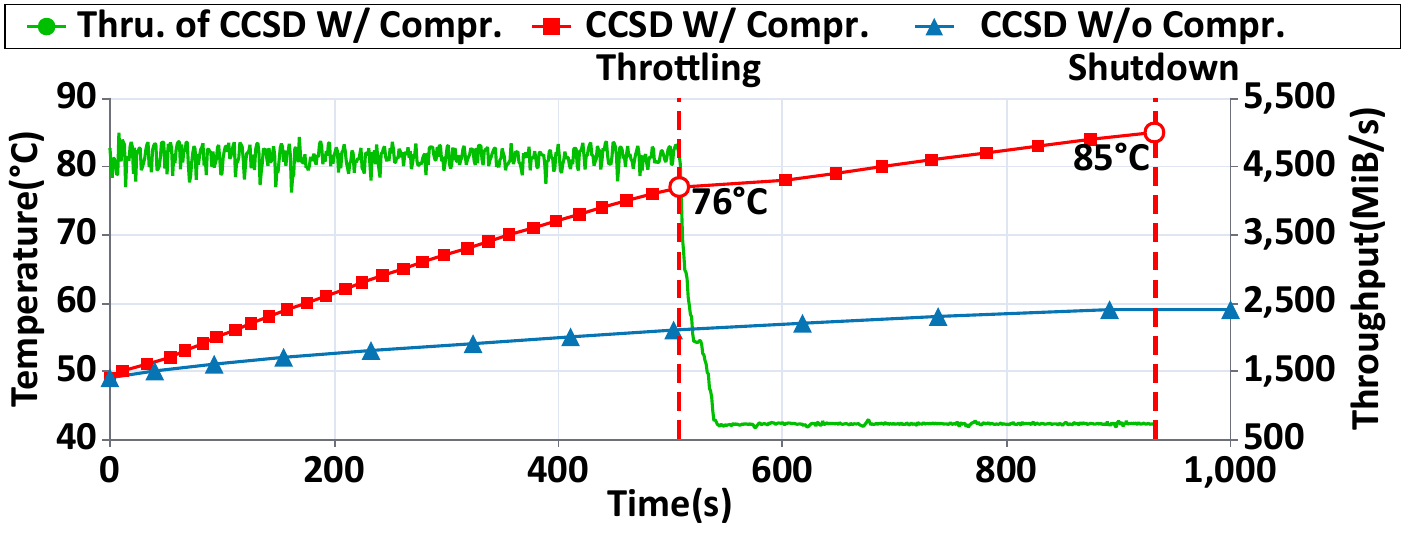}}
\caption{Temperature variation of CCSD with (W/, red line, Y1) and without (W/o, blue line, Y1) compression (Compr.). Thru. means throughput (green line, Y2).}
\label{fig:pow}
\end{figure}

Fig. \ref{fig:pow} shows the temperature variations over time\footnote{The temperature variation for decompression is shown in Fig. \ref{fig:tempCtl}.}.
When handling I/O activities without compression, the CCSD temperature increases slowly and saturates at 60$\degree$C. 
However, when compression is enabled, the temperature increases at a much faster pace (red line in the figure).
When the temperature reaches 76$\degree$C, the temperature increases at a slower pace because the in-CCSD overheating protection kicks in and throttles the throughput (green line) \cite{xu2019lessons}.
Unfortunately, the CCSD temperature  still increases despite a reduction in compression activities.
When the temperature reaches 86$\degree$C, the CCSD faces a thermal emergency and shuts down the device.

We then repeat the same experiment in a machine room with a cooling system (with the temperature set at 16$\degree$C). We simulate a server setting with four CCSDs (two ASIC-based and two FPGA-based) deployed in a machine rack. 
We observe the throughput throttling in 420s, which is only slightly better than that in Fig. \ref{fig:pow}.
The cooling system is ineffective as CCSD products currently lack in-package cooling, e.g., fans.
Alibaba, a major cloud service provider, also reported a similar result, i.e. the temperature of an idle SSD can be heated up to 78$\degree$C due to nearby working SSDs, when deployed in data centers with cooling \cite{xu2019lessons}.
Due to small factors, noise, cost, and marketing factors, it is yet clear if in-package cooling may be integrated into CCSDs shortly. 
A better solution is to proactively control CCSD temperature with software and hardware co-design.


\subsection{F2FS with Compression}\label{moti:B}
In contrast to CCSDs that support compression in hardware, F2FS integrates compression at the file system, which introduces software stack overhead and performance degradation. 
They are quantitatively evaluated as follows.

\subsubsection{Performance}\label{moti:performance}
We first compare the read and write performance when adopting Zstandard (ZSTD), LZ4, and LZO compression algorithms in F2FS.
Without loss of generality, we use four workloads representing random and sequential reads and writes, set the number of threads to 8, and set the compression ratio to 2 \cite{song2023f2fs}.
We choose the I/O size to be 128KB and 4KB for sequential reads/writes and random reads/writes, respectively.

\begin{figure}[htbp]
\centerline{\includegraphics[width=0.7\linewidth]{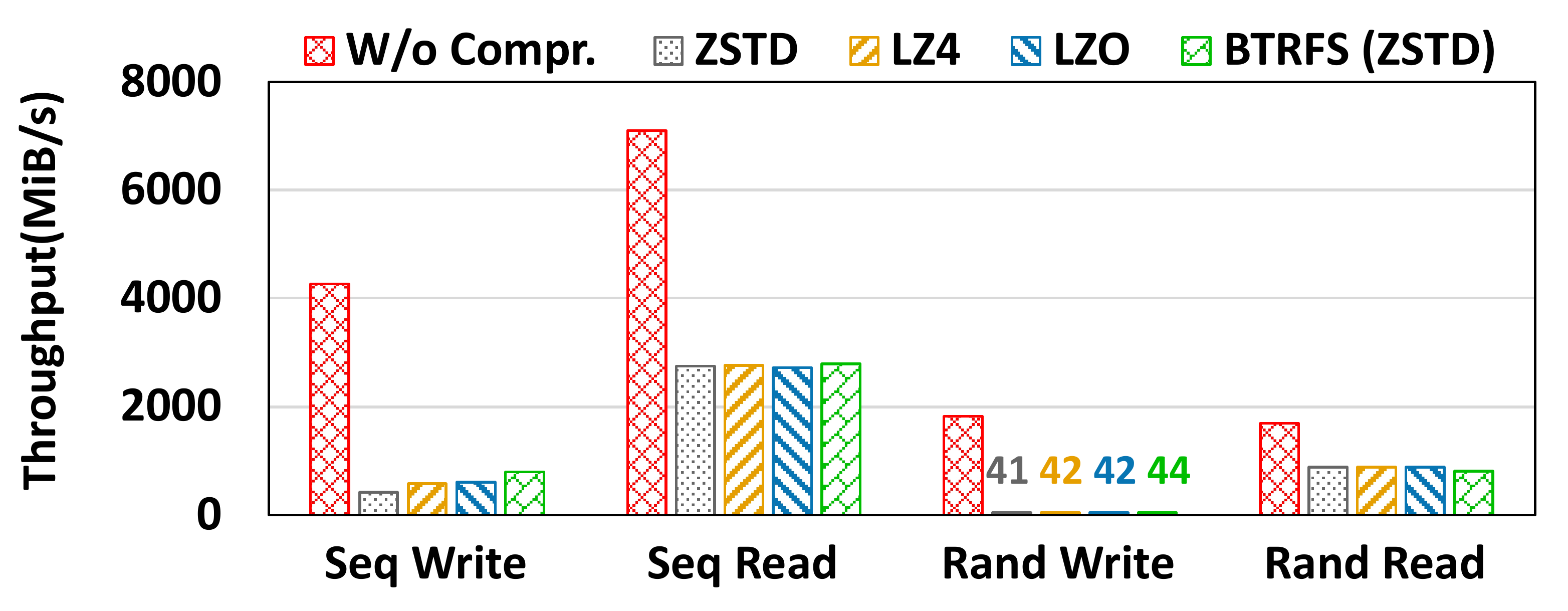}}
\caption{The throughput of F2FS with or without compression (W/o Compr.). The device side is CCSD with compression (before throttling).}
\label{fig:gc}
\end{figure}

Fig. \ref{fig:gc} summarizes the throughput of the four workloads.
From the figure, the throughput degrades under all workloads, due to (de)compression overhead.
On average, the throughput decreases by 87.6\% and 97.7\% for sequential and random write workloads, respectively, which is comparable to the 90\% performance degradation caused by temperature, as described in Fig. \ref{fig:pow}.
The difference between different compression algorithms is small, showing little impact on system throughput.
Also in the figure, read workloads exhibit a smaller throughput degradation than write workloads, i.e., 61.4\% and 48.0\% on average for sequential and random read workloads, respectively. 
This indicates that the impact of decompression is smaller than that of compression. 
The reason for this is that decompression is a straightforward process of rebuilding the original data based on a predefined set of instructions. 
In contrast, compression is a more complex task that requires traversing files and then performing compression calculations, such as determining offsets and encrypting, based on the chosen compression algorithm. 
The data is then sliced and added to specific locations within the compressed archive.
In addition, host-side decompression results in a smaller throughput reduction than the overheating-induced throttling described in \S~\ref{sec:te}.
We also evaluate BTRFS \cite{10.1145/2501620.2501623} with compression and observe similar results.
In summary, the throughput degradation from (de)compression on the host side is a common problem.

\subsubsection{CPU and Memory Overhead}
(De)Compression is also a computation-intensive task that consumes CPU and memory resources on the host side.
Next, we compare CPU usage and memory footprint with different compression algorithms.
The experimental setup is the same as above. 

\begin{figure}[htbp]
\centerline{\includegraphics[width=0.7\linewidth]{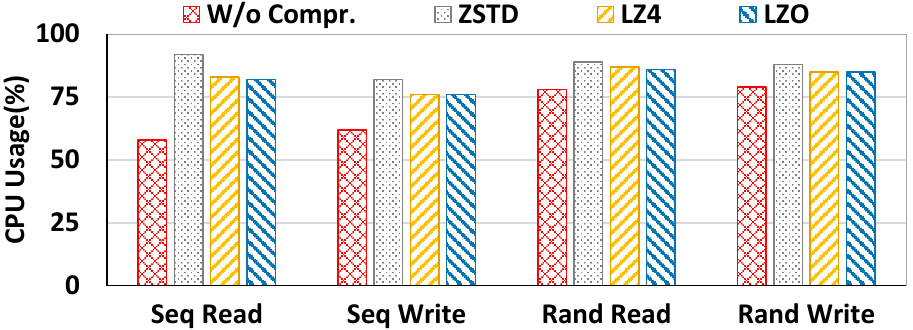}}
\caption{The CPU usage of F2FS with or without compression (W/o Compr.). The device side is CCSD with compression (before throttling).}
\label{fig:cpu}
\end{figure}

Fig. \ref{fig:cpu} compares the CPU usage of different compression algorithms.
The (de)compression operations require a significant portion of CPU cycles, i.e., the execution with LZ4 exhibits an average of 19.5\% CPU usage increase over the one without compression.
Different algorithms exhibit non-negligible differences in CPU utilization.
For example, the ZSTD compression requires more CPU cycles than LZ4 for sequential read and write workloads.
This aligns with the fact that LZ4 is a faster algorithm than ZSTD.
While not shown in the figure, the memory footprint for host-side compression is the size of the cluster times the number of threads, i.e., 16KB$\times$8=128KB.

\subsection{Takeaways}
We summarize our key observations as follows.
\begin{itemize}
\item \textbf{Observation-1}: The temperature of the CCSD should be well controlled to avoid thermal emergencies and service shutdowns.
A software-hardware co-design is preferred over a pure hardware solution.
\item \textbf{Observation-2}: On the host side, decompression operations have a smaller impact on system throughput than compression operations. 
The throughput degradation from host-side compression is comparable to the throttling strategy in CCSDs for avoiding thermal emergencies.
\item \textbf{Observation-3}: The (de)compression tasks consume CPU and memory resources on the host side.
By scheduling (de)compression tasks to CCSD, the resource consumption on the host side can be effectively reduced, minimizing the impact on co-running compute-intensive tasks.
\end{itemize}

\subsection{Current Compression-based Storage Systems}
Table~\ref {back1} summarizes existing compression-based storage systems.
FPC \cite{264858} proposed a pure host-side software compression approach that selects data for compression based on file types and employs a dual-mode foreground and background compression policy. 
In contrast, QZFS\cite{hu2019qzfs} integrates ASIC-accelerated compression within the ZFS file system, where it decides whether to compress based on the compression ratio and whether to offload the compression to the accelerator based on the source data size. 
Since neither of these schemes do not uses CCSDs, they fail to address the overheating problem.
COCO \cite{song2023f2fs} proposed to selectively compress data based on data type and compression ratio to reduce the temperature increase rate of CCSDs. 
Nevertheless, COCO’s effectiveness is limited when compressible data are abundant.
This paper proposes Waltz, which directly offloads compression tasks to the CCSD based on its temperature to effectively avoid overheating issues. 
Furthermore, this work is orthogonal to selective compression; the two approaches can be combined to further improve system performance and extend flash lifetime.

\begin{table}[hbp]
\caption{Comparison of compression-based storage systems}
\resizebox{1.0\linewidth}{!}{
    \begin{tabular}{c||c|c|c|c|c|c}
        \hline
        \hline
          & \textbf{CCSD} & \textbf{F2FS} & \textbf{FPC}\cite{264858} & \textbf{QZFS}\cite{hu2019qzfs}  & \textbf{COCO}\cite{song2023f2fs}& \textbf{Waltz} \\\hline
        \textbf{Software Compression}  & & \checkmark & \checkmark& \checkmark & & \checkmark\\
        \cline{1-7}
        \textbf{Hardware Compression} &\checkmark&&&\checkmark& \checkmark&\checkmark\\ 
        \cline{1-7}
        \textbf{Selective Compression Policy}& Always& Always& file types&  compressibility&\makecell{ data types and \\ compressibility}& Always\\
        \cline{1-7}
        \textbf{Compression Offloading Policy}& & & &source data size & & temperature\\
        \cline{1-7}
        \textbf{Overheating Prevention}& & & & & \checkmark&  \checkmark\\
        \hline
        \hline
    \end{tabular}
    \label{back1}
}
\end{table}

\section{Waltz: Temperature-aware Cooperative Compression}\label{sec:coco}
Waltz is a temperature-aware cooperative compression design to address the overheating problem, improve system performance, and prolong CCSD lifetime.

\subsection{Overview}\label{sec:over}
\begin{figure}[htbp]
\centerline{\includegraphics[width=0.7\linewidth]{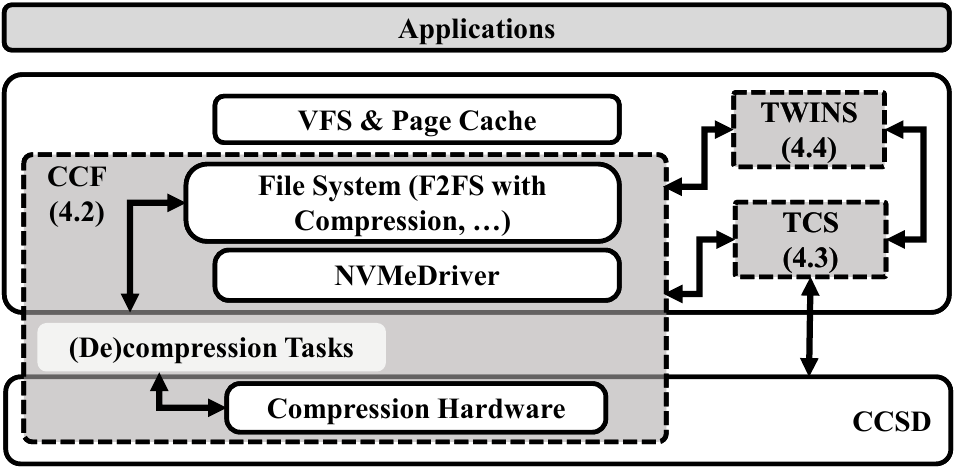}}
\caption{An overview of Waltz.}
\label{fig:over}
\end{figure}

Waltz schedules (de)compression tasks between the host side and the device side according to the system performance demands, as well as the temperature status of the CCSD. 
Fig.~\ref{fig:over} illustrates the overview of our Waltz design. 
It consists of three building blocks: cooperative compression framework (CCF), temperature-aware compression scheduling (TCS), and Waltz configuration selection (TWINS).
The CCF ensures that (de)compression tasks can be performed in cooperation between F2FS and CCSD (\S~\ref{sec:pacs}).
The TCS schedules the (de)compression tasks according to the temperature of the CCSD (\S~\ref{sec:tacs}).
The TWINS enables the selection of our optimization goal on either WAF or performance (\S~\ref{Waltzs and Waltzp}).


\subsection{Cooperative Compression Framework}\label{sec:pacs}
\begin{figure}[htbp]
\centerline{\includegraphics[width=0.7\linewidth]{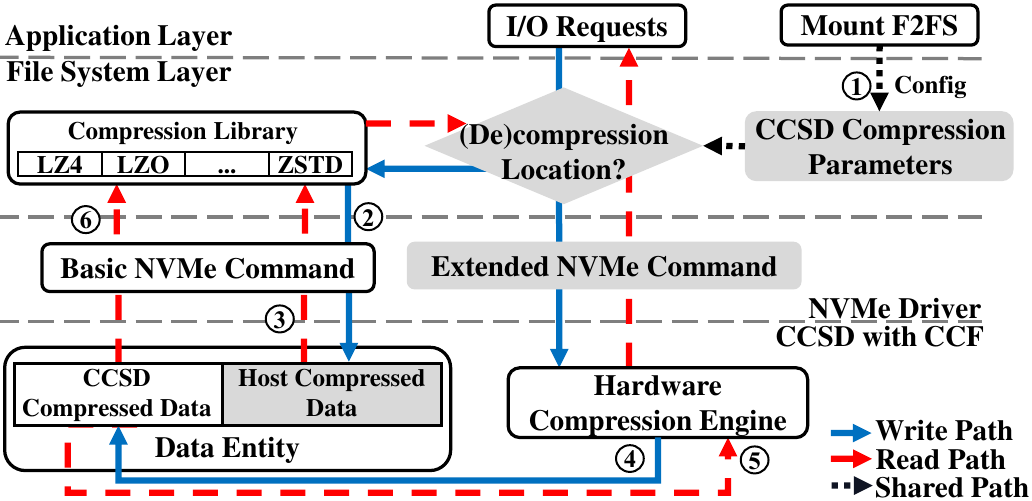}}
\caption{Workflow of CCF: \ding{174}, \ding{176}, and \ding{177} refer to decompression on the read path, and \ding{173} and \ding{175} refer to compression on the write path.}
\label{fig:des1}
\end{figure}

Mismatches between the host-side and device-side compression granularity, algorithms and metadata can lead to compressed data that cannot be properly decompressed.
The CCF component ensures that the compressed data, either on the host side or on the device side, can be properly decompressed. 
As shown in Fig. \ref{fig:des1}, the system initialization notifies the compression algorithm and the compression granularity to both F2FS and CCSD (\ding{172}).  
The I/O data, if compressed by F2FS, is written to the SSD directly (\ding{173}), which bypasses the hardware compression engine and thus avoids fruitless hardware compression. 
Such data can only be decompressed by F2FS and thus are always loaded to the file system before decompression (\ding{174}). 
While the I/O data, if compressed by CCSD (\ding{175}), can be decompressed by CCSD (\ding{176}) or F2FS (\ding{177}).

CCF, when servicing write requests, prioritizes CCSD compression unless it is temporarily paused by the TCS component.
CCF, when servicing read requests, exploits the metadata in F2FS to determine if the data are compressed by F2FS or CCSD.
For data compressed by CCSD, CCF may issue an extended NVMe read to enable hardware decompression or a normal NVMe read to fetch compressed data and then decompress in F2FS, depending if the TCS component temporarily pauses CCSD decompression.
For data compressed by F2FS, CCF does not exploit CCSD to decompress them.
This helps CCF cooperate with the encryption or other optimizations of F2FS. 
For example, if F2FS integrates a customized compression for special data, and adopts encryption during compression, the CCSD can bypass these data, leaving them to be handled by F2FS.
By scheduling (de)compression, CCF also avoids the waste of resources caused by duplicate (de)compression.

\subsection{Temperature-Aware Compression Scheduling} \label{sec:tacs}
The TCS component is responsible for determining if CCSD (de)compression tasks should be paused, depending on the current temperature of the CCSD.

\begin{figure}[htbp]
\centerline{\includegraphics[width=0.7\linewidth]{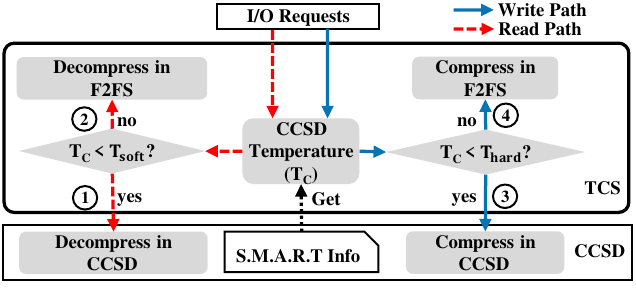}}
\caption{Workflow of TCS. The self-monitoring, analysis, and reporting technology information (S.M.A.R.T Info) contains the CCSD temperature.}
\label{fig:tacs}
\end{figure}

Fig.~\ref{fig:tacs} presents the TCS workflow. 
TCS sets two temperature thresholds $T_\text{soft}$ and $T_\text{hard}$, which is used to schedule decompression tasks and compression tasks, respectively.
$T_\text{soft}$ is smaller than $T_\text{hard}$ because we observe in \S~\ref{sec:obs} that decompression in F2FS has a smaller impact on system throughput.
Therefore, we can slightly trade off system throughput to reduce the CCSD temperature and achieve a more gradual change in system throughput.
The upper limit for $T_\text{hard}$ is $T_\text{emergency}$=86$^{\circ}$C,  which is the temperature that triggers a hardware thermal emergency and the system shutdown.
TCS periodically reads the CCSD temperature ($T_\text{C}$) using the in-package temperature sensor.
When $T_\text{C} < T_\text{soft}$, the system is free from overheating. 
TCS prioritizes CCSD (de)compression wherever possible in this case (\ding{172} and \ding{174}).

When $T_\text{soft} < T_\text{C} < T_\text{hard}$, the system enters \textit{temperature regulation phase}.
Decompression tasks are scheduled to F2FS (\ding{173}). 
Compression tasks are still scheduled to CCSD to minimize system throughput degradation.
When $T_\text{C} \ge T_\text{hard}$, the system enters \textit{temperature guard phase} as CCSD is about to trigger a thermal emergency.
TCS schedules compression tasks to F2FS (\ding{175}) as well, which leaves the computational engine in CCSD to be in the off state.
Normal I/O operations do not further increase the CCSD temperature, so that the CCSD temperature drops after scheduling all (de)compression tasks to F2FS.

\subsection{TWINS: Waltzs and Waltzp} \label{Waltzs and Waltzp}
Waltz provides a flexible cooperative compression framework that allows different optimization goals, focusing either on minimizing WAF or maximizing system throughput. 
The design that we discussed targets minimizing WAF as it enables data compression when possible. 
This is referred to as {\em Waltzs} in the paper.

Alternatively, Waltz can be configured to maximize system throughput. 
The scheme, referred to as {\em Waltzp}, minimizes the allocation of CPU cycles to (de)compression tasks. 
When $T_\text{soft} < T_\text{C} < T_\text{hard}$, Waltzp chooses CCSD for decompression so that the host throughput is not affected.
When $T_\text{C} \ge T_\text{hard}$, Waltzp disables data compression such that only decompression tasks might exist and be scheduled to F2FS.
Table~\ref{s and p} presents the two settings, Waltzs and Waltzp, and their operation decisions under different temperatures.
Waltzs and Waltzp can be mixed to meet the needs of different applications.
For example, Waltzp is used for latency-critical applications and Waltzs for other types of applications.


\begin{table}[htbp]
\small
\caption{Comparison of Waltzs and Waltzp. \label{s and p}}
    \begin{threeparttable}    
    \begin{tabular}{c||c|c|c}
        \hline
        \hline
          & $T_c < T_{soft}$& $T_c > T_{soft}$ & $T_c == T_{hard}$  \\
          \cline{1-4}
        \textbf{Waltzs}\tnote{*}  & \multirow{2}*{ CCSD (De)Compression} & \makecell[c]{CCSD Compression\\F2FS Decompression} & \makecell[c]{F2FS (De)Compression}\\
        \cline{1-1}
        \cline{3-4}
        \textbf{Waltzp}\tnote{*} &&\makecell[c]{CCSD (De)Compression}&\makecell[c]{F2FS Decompression}\\
        \hline
        \hline
    \end{tabular}
    \begin{tablenotes}
    \footnotesize
    \item[*] Waltzs is designed for space optimization, Waltzp is designed for performance optimization.
    \end{tablenotes}
    \end{threeparttable}
    
\end{table}

\subsection{Implementation and Analysis}\label{imp}
\subsubsection{Cooperative Compression Framework}\label{imp1}
We enhance F2FS to enable F2FS and CCSD integration. 
We allow flexible integration with three option flags.
The {\em -with\_CCSD=true} flag indicates if we mount the CCSD and enable the cooperative compression;
The {\em -granularity=4KB} flag specifies the granularity of the CCSD compression;
The {\em -algorithm=zlib} flag specifies the compression algorithm in CCSD.

To communicate if and how to compress data in SSDs, the host and the device communicate with requests packaged into a block I/O (bio) structure.
A reserved bit (e.g., bi\_flags) in bio is used as the flag bit.
If F2FS executes a (de)compression task, the flag bit is set to 1.
Otherwise, the flag bit is set to 0.
Due to the limitation of compression granularity, the data compressed by F2FS needs to be decompressed by F2FS.
The data compressed by F2FS can be identified by the metadata structure of F2FS.
Then, the flag bit is set to 1 when the data compressed by F2FS is accessed.
CCF has also been implemented to support filtering uncompressible data.
For uncompressible data (e.g., video files), neither F2FS nor CCSD should perform (de)compression tasks.
The highest bit of the logical address labels such data.

The CCSD must be informed whether to perform the (de)compression task.
The NVMe device driver generates NVMe commands based on the structure of the received bio request.
NVMe uses a 64-byte fixed-length command whose 8th to 15th bytes are reserved.
One bit of the 1st reserved byte in the NVMe command is used as the compression flag.
The compression flag is set to the same value as the flag bit in the bio structure.
When the NVMe command is sent to the CCSD, the CCSD determines whether to perform compression based on the compression flag.
The default value of the flag bit in the bio structure and the compression flag in the NVMe command is 0.
CCSD performs transparent (de)compression tasks when the compression flag of the received NVMe command is 0.
If the compression flag is 1, it skips the task.

\subsubsection{Temperature-Aware Compression Scheduler}
Modern CCSDs have integrated in-package temperature sensors and provided an access interface, e.g., the self-monitoring, analysis, and reporting technology (S.M.A.R.T.)~\cite{xu2019lessons} contains the CCSD's temperature that can be used by Waltz.
However, it is not only inefficient but also unnecessary to read the temperature for every I/O operation.
Therefore, we keep a kernel thread that reads the CCSD temperature from the S.M.A.R.T. information every second and makes it available during scheduling.
As the temperature changes relatively slowly, this frequency provides sufficient real-time information with low resource usage.
When servicing burst I/O requests that lead to fast temperature changes, Waltz may adjust the temperature reading frequency for better accuracy.

According to the CCSD's user manual and our empirical study (see Fig. \ref{fig:pow}), $T_\text{soft}$ and $T_\text{hard}$ are set to 76$\degree$C and 85$\degree$C, respectively. 
$T_\text{hard}$ is close to $T_\text{emergency}$ and has small room to be further increased.
In this paper, we set $T_\text{soft}$ to the throttling-triggered temperature for a fair comparison.
In real scenarios, $T_\text{soft}$ can be set at a high temperature to fully exploit the performance benefits of CCSD.
For example, to service read-intensive or computation-intensive workloads, Waltz may further optimize performance and resource consumption by turning up $T_{soft}$.
Waltz's co-design can fully utilize the semantic information on the host side to adjust $T_{soft}$.
We will explore dynamic adjustments to $T_{soft}$ in the future.

\subsubsection{Overhead Analysis}
We analyze the overhead in the following three aspects.

\noindent
\textbf{Memory overhead:}
Waltz introduces several variables in CCF to synchronize F2FS and CCSD. 
The space overhead is three bytes.
It also introduces two reserved bits in the bio structure and NVMe command. 
It exploits the highest bit in the logical address to minimize the overhead.
Waltz also introduces three variables in TCS to record the temperature and thresholds. 
Its space overhead is also three bytes.
In summary, the space overhead is negligible.

\noindent
\textbf{Time overhead:}
Waltz needs to test the compression flag bit in CCF, which is negligible.
Waltz also needs to fetch the CCSD temperature in TCS, which is not on the critical path.
Its impact on system performance is also negligible.

\noindent
\textbf{Other overheads:}
Waltz is implemented as a configurable model, which can be optionally enabled in F2FS.
Waltz takes the least intrusive design in F2FS such that, even if Waltz is active, selected streams of data can be compressed and/or encrypted, without interference with Waltz.

In summary, Waltz is a low-cost cooperative compression scheme that is ready to be deployed in modern CCSDs.

\section{Case Study: Utilizing Compressed Space for Reserved Space}\label{sec:case}

\subsection{Reserved Space Size for F2FS}\label{motivation:C}
Data compression leads to reduced SSD space occupation, with the saved SSD space traditionally being exploited to amortize flash writes and thus extend the SSD lifetime. 
In this section, we present cooperative utilization of the saved space for a better tradeoff between SSD lifetime and F2FS performance.

\begin{figure}[htbp]
\centerline{\includegraphics[width=0.7\linewidth]{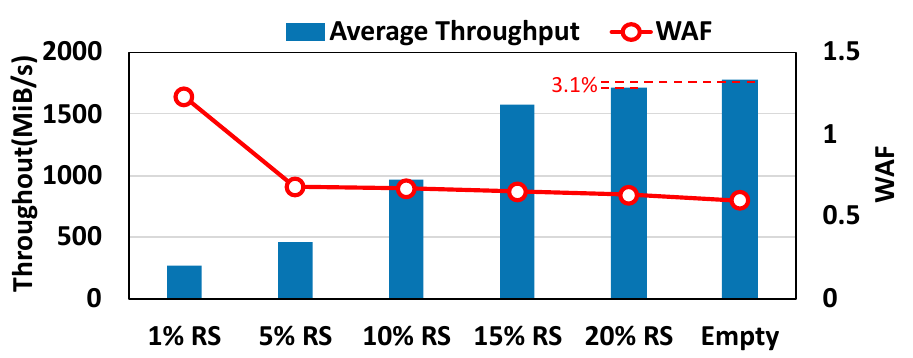}}
\caption{The WAF and throughput of F2FS with CCSD at different RS sizes (before throttling). 
}
\label{fig:waf}
\end{figure}

F2FS is an LFS (log-structured file system) that supports segment cleaning (SC) by allocating a reserved space (RS) of a fixed percentage, e.g., 1\%, of the space. 
We conduct an experiment to study the correlation between the F2FS performance and the size of the RS and summarize the experimental results in Fig. \ref{fig:waf}. 
In the experiment, the compression ratio of the data is set to 2, and we allocate certain percentages of SSD space as RS space and evaluate the F2FS system throughput and WAF. 
We fill the partition of F2FS to trigger the SC and then run the evaluation.

With increasing RS space, the WAF factor decreases while the throughput improves.
This is because a large RS space helps to consolidate writes and thus reduces WAF, which improves the throughput.
However, the benefits diminish when the RS space increases beyond 20\% of the space. 
For example, only a 3.1\% difference for throughput between allocation 20\%S and the {\tt empty} scheme, i.e., the SSD is nearly empty so that I/O writes won't trigger SC.


\subsection{On-Demand Space Allocation}\label{sec:odsa}
Although F2FS can proactively allocate a large percentage of free space as RS space, it is less preferable due to reduced usable logical space in the file system.
Instead, we develop a cooperative allocation that adjusts the size of the RS space according to the compression ratio of the data saved in the SSD.

Intuitively, saving logical data in compressed format in an SSD occupied a smaller SSD space than that in uncompressed format.
Given an SSD with capacity $S$, both SSD and F2FS have an address space of size $S$ if storing uncompressed data.
Assuming the compression ratio is 2, saving $S$ logic data in the compressed format results in $S$/2 occupied SSD space, leaving the other $S$/2 or 50\%$S$ space unused.
The unused SSD space can be allocated as RS space for performance improvement.
Given the compression ratio changes at runtime, the RS space cannot occupy more than the space saved from compression. 
We shall not allocate more than 20\%$S$ as the RS space according to the observation in Fig. \ref{fig:waf}.

\begin{figure}[htbp]
\centerline{\includegraphics[width=0.7\linewidth]{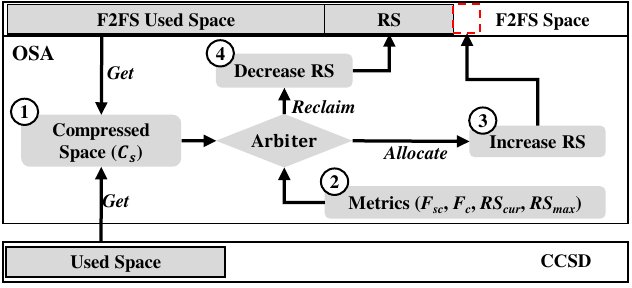}}
\caption{Workflow of OSA. The arbiter adjusts the RS based on the metrics.}
\label{fig:des2}
\end{figure}

\textbf{Space allocation and reclamation:}
Fig. \ref{fig:des2} shows the workflow of our on-demand RS allocation (OSA) scheme.
The OSA scheme decides (\ding{172}) if the current RS space needs to be expanded or reduced based on the invocation frequency of the SC and the space saved from compression. 
The decision is made based on the following parameters:
\begin{itemize}
\item ${C_s}$: the size of the space saved from compression, i.e., the difference between F2FS allocation and its occupied SSD space;
\item $RS_{cur}$: the size of the extra RS space, i.e., this is in addition to the space allocated by F2FS.
\item $RS_{max}$: the threshold of the maximal extra RS space;
\item $T_{s}$: the unit of the RS space for expansion or shrinking;
\item $F_{sc}$ and $F_{c}$: the invocation frequencies of SC cleaning and data copying in SCs, respectively.
\item $FT_{sc}$ and $FT_{c}$: the invocation threshold of SC cleaning and data copying, respectively.
\end{itemize}

We decide to expand the RS space (\ding{173}) by $T_{s}$ if ($F_{sc} > FT_{sc}$ or $F_{C} > FT_{C}$), $RS_{cur}+T_s < RS_{max}$, and $RS_{cur}+T_s < C_s$;
we decide to shrink the RS space (\ding{174}) by $T_{s}$ if ($F_{sc} < FT_{sc}/2$ and $F_{C} < FT_{C}/2$), or $RS_{cur}+T_s > C_s$;
Otherwise, we keep the RS space unchanged. 
We ensure $RS_{cur} \ge 0$ during shrinking.
Based on our offline study, we set $T_{sc}$, $T_{copy}$, $T_s$, and $RS_{max}$ to 4 times per second, 256 times per second, 1\% of the space, and 20\% of the space, respectively. 
We will detail the study in \S~\ref{sec:eva}.

\textbf{Implementation and overhead analysis.}
We implemented the OSA scheme as a kernel thread that manages the RS space, without affecting the SC of F2FS.
It is triggered once per second and collects the parameters as listed above to allocate and reclaim the space saved by compression. 
The size of the space saved by CCSD compression can be extracted by exploiting the interface provided by the CCSD. 
The flexible settings are achieved by extending mounting options, with the details discussed in \S~\ref{sec:setup}.
The time overhead of OSA is on the adjustment of the RS size, which is O(1).

\textbf{Discussion.} 
There are several approaches to exploit the SSD space saved from data compression.
The most traditional approach is to amortize the flash writes from the host and thus extend the SSD lifetime.
This is the most popular one as it can be naturally integrated into the existing systems.
Alternatively, the saved space can be exposed to F2FS such that F2FS addresses a larger address space.
This is less popular because the compression ratio may change at runtime, complicating the management of potential mismatches between the size of the physical space and that of the logical space visible to the user.
While the saved space can be made available as the RS of F2FS and the OP space of CCSD directly, a mismatch is still possible as RS and OP are not visible to users.

In this paper, we adopt the first approach as the baseline. 
We allocate a portion of the saved space as the RS space while leaving the rest for extending the SSD lifetime, the same as that in the baseline.
We will explore more ways in the future.



\section{Evaluation}\label{sec:eva}
We evaluate Waltz to answer the following questions:
\begin{itemize}
\item How is the effect of the temperature control of Waltz with different access patterns? (\S~\ref{sec:tc})
\item How is the performance of Waltz compared with that of state-of-the-art methods? (\S~\ref{sec:per1} and \ref{sec:per2})
\item How does the compression ratio of workloads affect performance? (\S~\ref{sec:cr})
\item How does Waltz perform under real workloads? (\S~\ref{sec:rw})
\item How does Waltz perform in terms of WAF? (\S~\ref{sec:waf})
\item How does Waltz perform in terms of CPU usage and memory overhead? (\S~\ref{sec:cpu})
\end{itemize}

\subsection{Experimental Setup}\label{sec:setup}

\textbf{Testbed:}
Table \ref{tab:testbed} shows the detailed configuration of the testbed.
The partition size of F2FS is 200GB, and the compression algorithm for F2FS is ZSTD.
The host CPU has 16 physical cores and 32 hyperthreads.
Implementing the cooperative compression framework of Waltz necessitates modifications to the CCSD firmware to determine which data to compress on the device. 
Because the real implementation of a CCSD platform involves hacking the device controller and revising the protocol, which is a complex and often infeasible task, we circumvented this challenge by first building a temperature model from real ASIC-based CCSDs. 
Subsequently, we collected their performance characteristics under different temperature states to guide the configuration of our emulator (FEMU \cite{li2018case}).
We use the performance of the setting with the compression ratio being 1 to represent the performance of CCSD without compression, and the performance just before throttling to represent the performance without throttling.
The performance of F2FS with compression is tested before throttling is triggered on the CCSD product.
We then integrate the fitted temperature modeling into the emulator, and take the I/O traces generated by the workload for evaluation.

\begin{table}[t]
\caption{Testbed Configurations and Real Workloads Characteristics\label{tab:testbed}}
\begin{center}
\resizebox{0.7\linewidth}{!}{
\small
\begin{tabular}{c|c|c|c|c}
\hline
\hline
\multirow{4}*{\textbf{Host}} & CPU & \makecell[c]{AMD Ryzen R9-7950X\\ @ 5.7GHz (16C32T)}&\multirow{2}*{\textbf{Workloads}} & \multirow{2}*{\textbf{\makecell[c]{Compression \\ Ratio}}} \\
\cline{2-3}
& Memory & 32GB& & \\
\cline{2-5}
& OS & Ubuntu 22.04& Webserver & 1 - 2\\
\cline{2-5}
& Kernel & Linux 5.19.0& Varmail & 2 - 3\\
\hline
\multirow{2}*{\textbf{Device}} & Type & CCSD & Fileserver & 2 - 3\\
\cline{2-5} 
& Storage & 3TB & OLTP & 3 - 4\\
\hline
\hline
\end{tabular}
}
\end{center}
\end{table}

\begin{figure}[htbp]
\centerline{\includegraphics[width=0.7\linewidth]{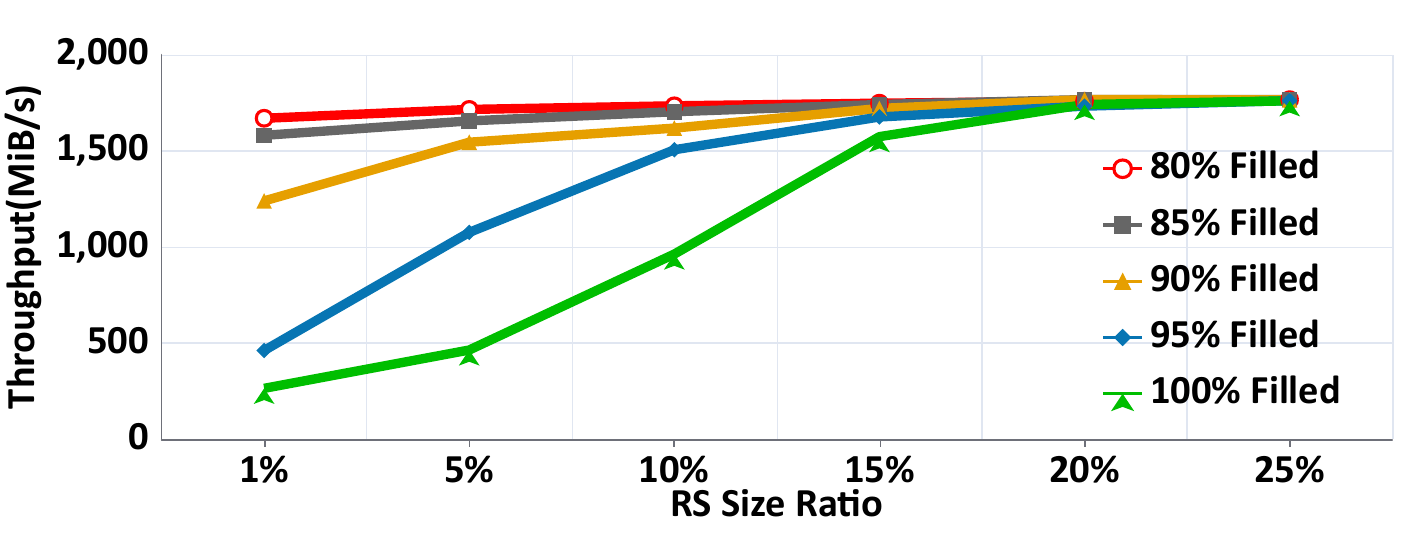}}
\caption{The effect of the RS size on performance (before throttling).}
\label{fig:sen}
\end{figure}

\textbf{Workloads:} 
We evaluate Waltzs and Waltzp using both microbenchmarks and real workloads.
FIO \cite{fio} is the microbenchmark used to generate various synthetic workloads.
Without loss of generality, the compression ratio is set to 2 by default, and the I/O sizes are set to 128KB and 4KB for sequential and random access workloads, respectively.
All tests are run for 20 minutes with the temperature control (i.e., throttling or co-compression) triggered.
We repeated each test case three times and reported the average. 
As to real workloads, we choose four typical ones from Filebench \cite{filebench}.
Table \ref{tab:testbed} shows their characteristics. 
The compression ratio is determined by the types of these workloads. 
In the case of OLTP, the compression ratio is usually between 3 and 4 \cite{zuck2014compression}.
Without loss of generality, we assume that the compression ratio of real workloads follows the Zipfian distribution \cite{green1986pareto}.
\\
\textbf{Configuration:}
We use an offline evaluation to find \enquote{rsr\_max}.
Fig. \ref{fig:sen} shows the throughput as the percentage of the RS varies from 1\% to 25\%.
Different curves correspond to different degrees by which the partition is filled, varying from 80\% to 100\%.
These curves show that the performance peaks when the RS size ratio is 20\%.
\enquote{rsr\_max} mainly acts as a limitation, so we should consider the worst-case scenario (100\% filled) to set it.
Based on this, we set \enquote{rsr\_max} to 20\% to simplify the evaluation.
The partition size and the performance of the storage device may affect \enquote{rsr\_max}.
We recommend setting \enquote{rsr\_max} according to actual situations.
\\
\textbf{Compared methods:}
To demonstrate the effectiveness of \textbf{Waltz}, four additional schemes are evaluated.

\begin{itemize}
\item \textbf{Baseline.} 
It represents pure CCSD compression and utilizes the throttling mechanism to prevent overheating. 
Throttling mitigates the temperature increase with reduced throughput. 
Unfortunately, the temperature of the CCSD may still increase with reduced compression activities, which can lead to a thermal emergency, as shown in \S~\ref{sec:te}.
\item \textbf{F2FSC.} It represents F2FS with compression running on CCSD without compression. 
In this scheme, the CCSD throttling is not triggered and there is no overheating problem.
\item \textbf{FPC.} It represents selective compression on F2FS \cite{264858} or CCSD \cite{song2023f2fs}.
In FPC, the data selected for compression are based on a hybrid metric of the file types \cite{264858} and access frequency \cite{song2023f2fs}
\footnote{The differences to \cite{264858,song2023f2fs} also include: (1) \cite{264858} does not consider CCSD compression while \cite{song2023f2fs} does not consider F2FS compression; (2) FPC uses the latest CCSD product; (3) FPC uses a fixed ambient temperature.}, but not temperature states.
FPC compresses selected data in F2FS and other data in CCSD.
The CCSD may still be overheated without temperature control.
\item \textbf{TCS.} It represents the proposed cooperative compression but does not include the OSA scheme.
\item \textbf{Waltzs.} It represents the proposed cooperative compression that focuses on optimizing WAF.
\item \textbf{Waltzp.} It represents the proposed cooperative compression that focuses on optimizing system throughput.

\end{itemize}

\subsection{Temperature Control}\label{sec:tc}

\begin{figure}[htbp]
\centerline{\includegraphics[width=0.7\linewidth]{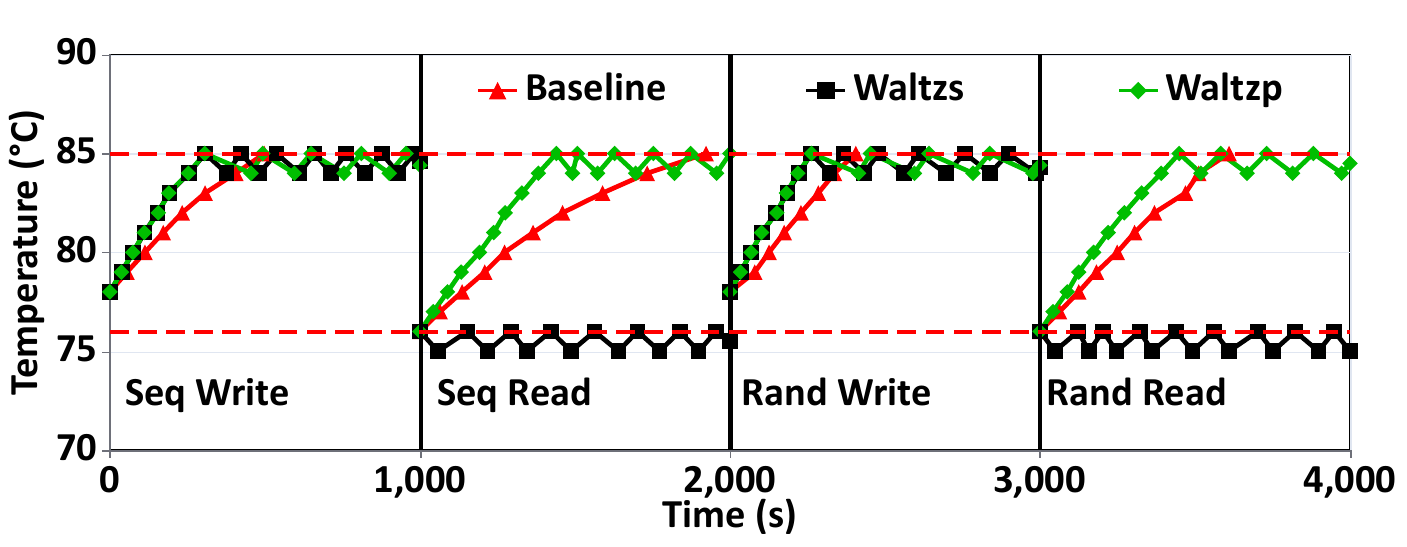}}
\caption{Temperature variation in four access patterns. 
Waltzs and Waltzp are designed to avoid the overheating problem of CCSD.
}
\label{fig:tempCtl}
\end{figure}

Given one of our major design goals is to control CCSD temperature to avoid overheating, we first analyze different temperature control decisions.
Fig. \ref{fig:tempCtl} summarizes the temperature changes in four types of workloads.
We re-initialize the CCSD to ambient temperature at the beginning of processing each type of workload. 
The temperatures of Baseline are read from the in-package temperature sensor, while those of Waltzs and Waltzp are derived from the model in the emulator.
From the figure, Baseline encounters the overheating problem and shuts down the SSD under all workloads. 
For example, the CCSD gets overheated in 582 seconds under a sequential write workload.  

Waltzs and Waltzp are effective in controlling the SSD temperatures --- neither is CCSD overheated nor is it shut down with these schemes.
The temperatures of Waltzs are not higher than T$_\text{soft}$=76$\degree$C when processing read I/Os.
This is because Waltzs moves decompression activities to F2FS when the CCSD temperature is higher than T$_\text{soft}$. 
Since there are no compression activities in read-only I/Os, the CCSD becomes a normal SSD such that its temperature cannot increase more than T$_\text{soft}$.
Waltzp continuously uses CCSD to decompress and thus the CCSD temperature may hit $_\text{hard}$.
The CCSD temperatures can hit T$_\text{hard}$ for either scheme with write I/Os.

\begin{figure}[!thp]
\centerline{\includegraphics[width=0.7\linewidth]{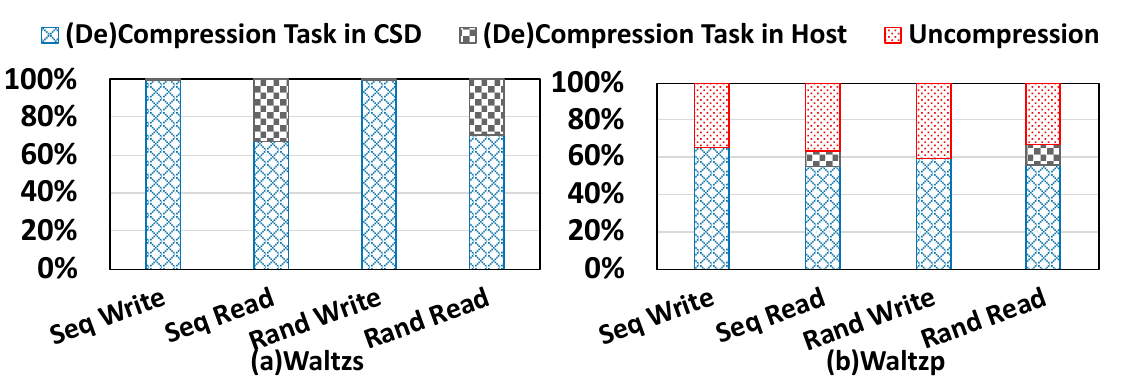}}
\caption{Distribution of (de)compression tasks on the host- and device-side.}
\label{fig:exp07}
\end{figure}

To study the effectiveness of scheduling in Waltz, we then quantify the distribution of (de)compression tasks on both sides and summarize the results in Fig. \ref{fig:exp07}.  
The distributions are calculated based on the ratios of temperature rise and fall times and the throughput of the host side and device side on the CCSD product.

From the figure, CCSD performs the majority of the (de)compression activities for all four types of workload. 
On average, only 16.0\% of (de)compression tasks are scheduled on the host side.
In particular, for write-only I/Os, on average, 99.3\% of compression tasks are scheduled on the device side. 
This is because compression tasks are computation-intensive. 
Once the CCSD temperature reaches T$_\text{hard}$, Waltzs schedules compression tasks to the host side, which brings large throughput degradation and cools the CCSD effectively. 
The Waltzs scheme can then schedule incoming compression tasks to the CCSD.

As a comparison, Waltzp disables compression when the CCSD temperature reaches T$_\text{hard}$.
While Waltzp helps to minimize the impact on the system performance, it misses 34.7\% compression opportunities on average.

\subsection{Microbenchmark}
Next, we study the system throughput under different schemes.
To understand the effect of the OSA scheme, we evaluate the throughput in two settings, i.e., with and without SC triggered, respectively.

\subsubsection{Without SC}\label{sec:per1}
\begin{figure}[htbp]
\centerline{\includegraphics[width=0.7\linewidth]{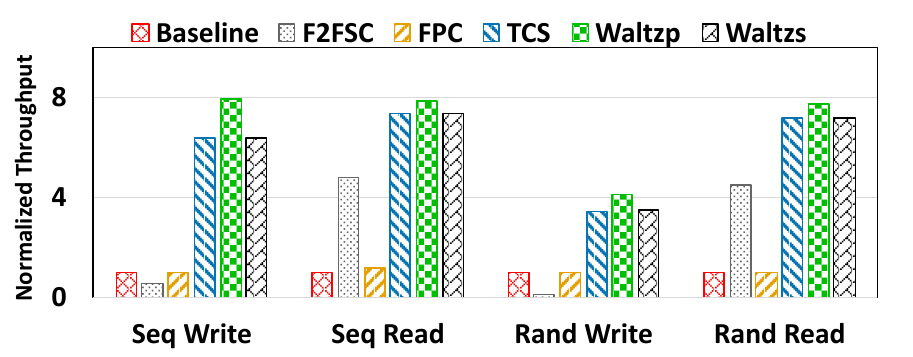}}
\caption{Normalized average throughput \textbf{without SC}.}
\label{fig:exp01a}
\end{figure}

Fig. \ref{fig:exp01a} compares the throughput of different schemes when SC is disabled.
All results are normalized to the Baseline. 
From the figure, F2FSC exhibits degraded write throughput and greatly improved read throughput over the Baseline, i.e., on average 67\% write throughput degradation and 367\% read throughput improvement, respectively. 
This is because the throughput degradation from host-side compression is more severe than the throughput drop caused by CCSD throttling, resulting in lower write bandwidth compared to the Baseline. 
In contrast, read bandwidth improves because host-side decompression has a smaller impact on system performance.

FPC is not temperature-aware and thus inevitably triggers throttling. FPC achieves a similar performance compared to that of the Baseline. 
The temperature-aware TCS achieves large throughput improvements by eliminating throttling and co-compression. 
On average, the write throughput of TCS improves by 391.5\%, while the read throughput improves by 627\% over the Baseline.
Waltzs has the same throughput as the TCS because no SC is triggered.
Waltzp achieves the best improvements by avoiding host-side compression overhead and aggressive scheduling (based on $T_{hard}$).
On average, the throughput of Waltzp is 13.9\% higher than that of Waltzs.

\subsubsection{With SC}\label{sec:per2}
\begin{figure}[hbp]
\centerline{\includegraphics[width=0.7\linewidth]{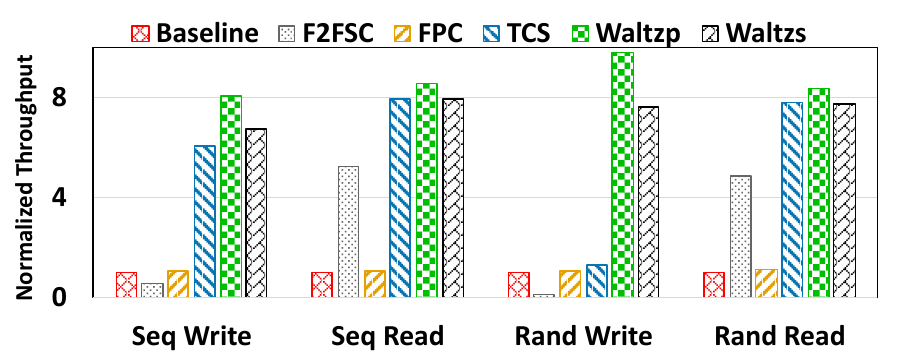}}
\caption{Normalized average throughput \textbf{with SC}.}
\label{fig:exp01b}
\end{figure}

Fig. \ref{fig:exp01b} shows the normalized throughputs of different schemes when the SC is triggered. 
During the evaluation, we fill in the F2FS to trigger the SC and then run the throughput evaluation. 

Waltzp achieves larger throughput improvements compared to the setting without SC.
Since SC operations block user requests and degrade the system performance, the write throughput of F2FSC decreases by 65.5\% while the read throughput improves by 407.5\% over the Baseline.
In addition, FPC ignores the space saved by compression and thus fails to reduce the SC overhead.
FPC achieves a similar read throughput as that of the Baseline.

Compared to the Baseline, the write throughput of TCS is improved by 270\%, and the read throughput is improved by 689.5\% on average.
Waltz improves the throughput by allocating part of the space saved by compression as the RS space. 
The write throughput of Waltzs is improved by 95.5\% compared to that of TCS.
Compared to Waltzs, the throughput of Waltzp is improved by 23.6\%.
These results prove that the OSA scheme effectively utilizes the space saved by compression to optimize SC.

\subsubsection{Sensitive Study on Compression Ratio}\label{sec:cr}
\begin{figure}[!htp]
\centerline{\includegraphics[width=0.7\linewidth]{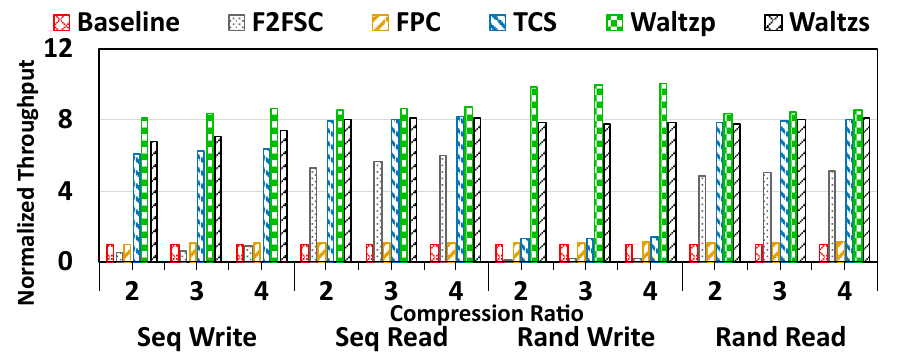}}
\caption{Normalized average throughput by varying compression ratios \textbf{with SC}.}
\label{fig:exp06}
\end{figure}

To understand the impact of compression ratio on Waltz, we study the system throughput with different compression ratios.
Fig. \ref{fig:exp06} shows the normalized average throughput by varying the compression ratios (i.e., 2, 3, and 4).
The \enquote{x} in \enquote{Baseline-x} represents the compression ratio.
First, when the compression ratio increases, the performance of all schemes is improved.
Second, throttling after triggering temperature control limits the effect of compression ratio on performance.
Compared to FPC-2, FPC-4 improves write and read performance by 9.8\% and 5.7\% on average, respectively.
Third, as the compression ratio increases, the amount of data actually written will decrease, thus improving performance.
Compared to F2FSC-2, F2FSC-4 improves write and read performance by an average of 36.3\% and 9.5\%, respectively.
Then, compared to Waltzs-2, Waltzs-4 improves write performance by 4.2\% and read performance by 3.1\% on average.
Finally, compared to Waltzp-2, Waltzp-4 improves write performance by 4.6\% and read performance by 1.3\% on average.
In conclusion, the proposed design is effective with different compression ratios.

\subsection{Real-World Workloads}
We also evaluated Waltz using real workloads to understand its performance and WAF.
The real workloads are more complex compared to Microbenchmarks.

\subsubsection{Performance}\label{sec:rw}
\begin{figure}[h]
\centerline{\includegraphics[width=0.7\linewidth]{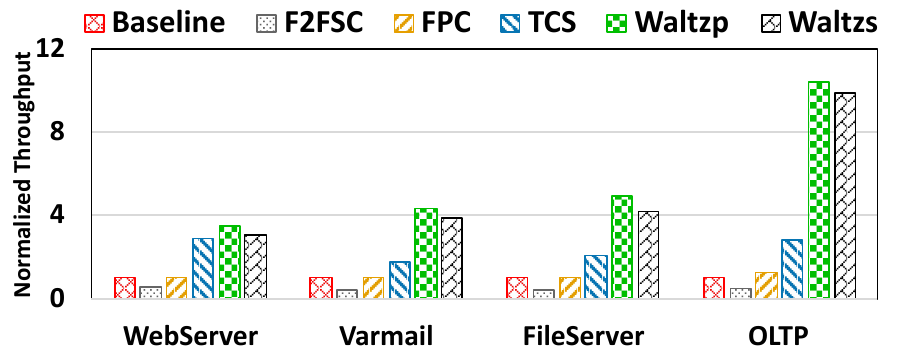}}
\caption{Normalized average throughput \textbf{with SC}.}
\label{fig:exp03}
\end{figure}

Fig. \ref{fig:exp03} shows the results.
First, due to the compression overhead, F2FSC has the worst performance.
The performance of F2FSC drops by 52\% on average compared to the Baseline.
Second, compared to F2FSC, the performance of FPC is improved by 126.6\% on average.
Third, TCS schedules (de)compression tasks based on temperature, which improves the performance.
Compared to FPC, the performance of TCS improves by 120.9\% on average.
Finally, by using some of the space saved by compression to increase RS, Waltz reduces the SC overhead of F2FS.
Waltzs improves its performance by an average of 118.3\% compared to TCS.
Waltzp avoids the performance degradation caused by host-side compression.
The performance of Waltzp is further improved by 10.3\% on average compared to Waltzs.
These results show that Waltz achieves optimal performance when serving complex real-world workloads.

\subsubsection{WAF}\label{sec:waf}
\begin{figure}[htbp]
\centerline{\includegraphics[width=0.7\linewidth]{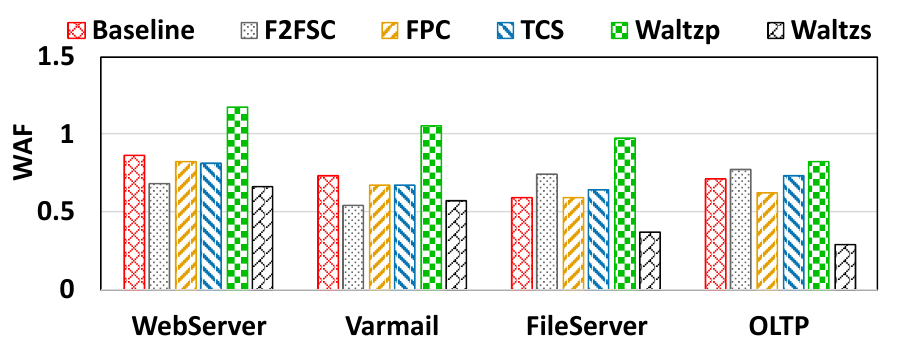}}
\caption{WAF under real-world workloads \textbf{with SC}.}
\label{fig:exp04}
\end{figure}

We use the ratio of the amount of data written on the host side (before compression) to the amount of data written on the flash memory to calculate the WAF. 
Fig. \ref{fig:exp04} shows the WAF comparison under six schemes.
In the results, WAF is lower than 1 because the total amount of data written to storage is lower than that of the host.
First, F2FSC uses 16KB compression granularity by default, which enlarges 4x compared to the Baseline's 4KB compression granularity.
Larger compression granularity typically results in a smaller compressed size.
However, F2FSC updates any part of the 16KB data, which requires decompressing, updating, compressing, and then writing back, thus increasing the WAF.
Compared to Baseline, F2FSC decreases WAF by 23.3\% under sequential access patterns but increases the WAF by 16.2\% in random access patterns.
Second, since most of the data has been compressed by CCSD, FPC achieves a WAF similar to that of the Baseline.
Third, TCS schedules the execution of (de)compression tasks based on temperature, which cannot optimize WAF.
Thus, the WAF for TCS is between the Baseline and F2FSC.
Finally, Waltzp is optimized for performance.
Uncompressed data will lead to a high WAF as a side effect of optimal performance.
Waltzp's WAF deteriorated by 38.8\% on average compared to the Baseline.
Waltzs achieves the optimal WAF among all methods.
It always expects compression compared to Waltzp.
By reducing the SC overhead, Waltzs's WAF is reduced by 34.6\% on average compared to Baseline.
With these results, both Waltzs and Waltzp achieve their optimization objectives.

\subsubsection{CPU Usage and Memory Overhead}\label{sec:cpu}
\begin{figure}[htbp]
\centerline{\includegraphics[width=0.8\linewidth]{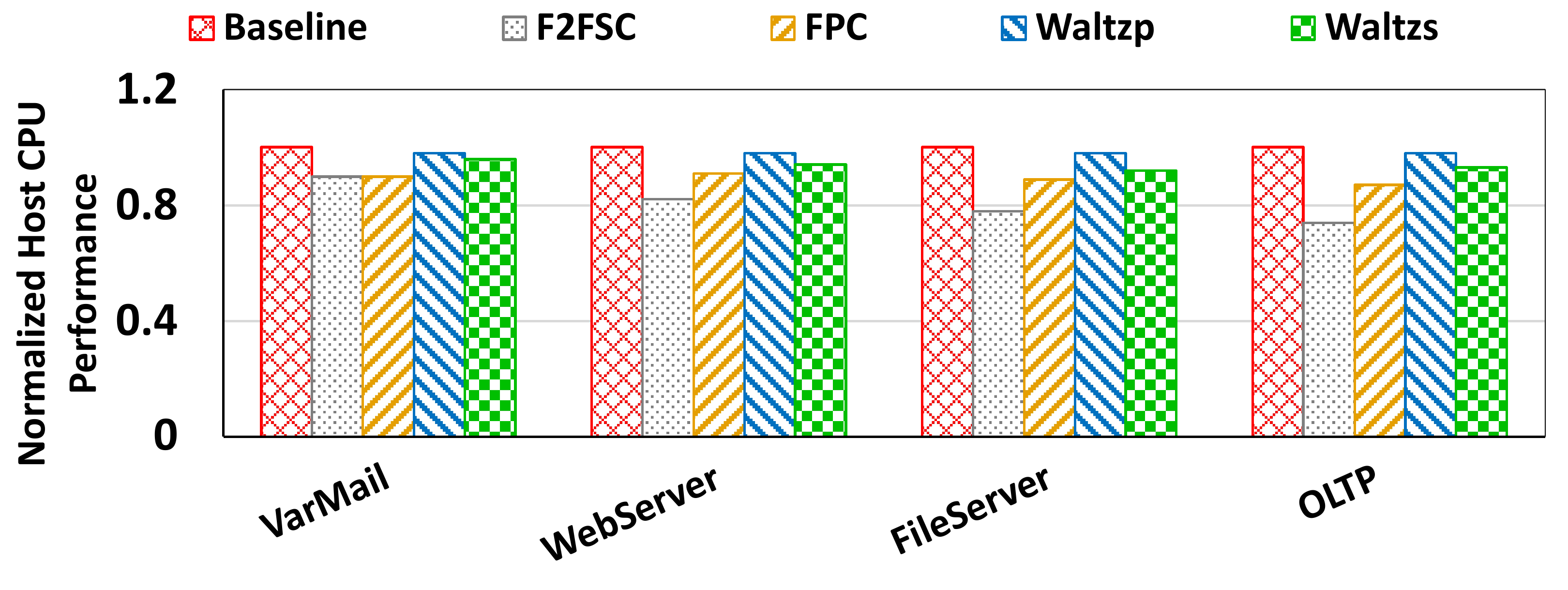}}
\caption{The performance of computation-intensive workloads under running both computation-intensive and I/O-intensive workloads.}
\label{fig:exp05}
\end{figure}

After demonstrating Waltz's superior performance, we turn to evaluating Waltz's host-side CPU usage and memory overhead.
Sysbench \cite{sysbench} was used to evaluate CPU performance.
The CPU performance is quantified to the cycles CPU looped to calculate the prime numbers from 1 to 10,000,000 in 20 minutes.
We evaluated running computation-intensive and I/O-intensive workloads together to collect CPU cycles for the computation-intensive workloads.
Fig. \ref{fig:exp05} shows the average results of 10 evaluations.

First, we can see that computation-intensive workloads achieve optimal performance under the Baseline.
This is because CCSD handles all (de)compression tasks, thus avoiding consuming host-side resources.
Second, F2FSC has the highest CPU usage due to the fact that the (de)compression tasks are performed on the host side.
Compared to the Baseline, the performance of compute-intensive workloads under F2FSC drops by 19.3\% on average.
Third, FPC reduces CPU usage by selective compression on the host side.
The performance of computation-intensive workloads under FPC is improved by an average of 10.2\% compared to F2FSC.
Finally, Waltz (Waltzs and Waltzp) always expects to schedule (de)compression tasks to the CCSD without triggering the overheating problem, thus minimizing the resource consumption on the host side.
The performance of computation-intensive workloads under Waltz is improved by an average of 17.9\% compared to F2FSC (only 4.1\% lower than Baseline).
Waltzp further reduces CPU usage by disabling host-side compression.
Furthermore, compared to F2FSC, Waltz reduces the memory overhead of host-side compression by 88.7\%.
These results show that Waltz can effectively reduce resource contention on the host side by scheduling (de)compression tasks with sufficient consideration of the CCSD's advantages.

\section{Related Work}\label{sec:related}

\textbf{Host-side compression.}
Compression is an effective way to reduce the amount of data written to NAND flash.
Some studies implement compression on the host side \cite{cF2FS,e2compr}.
A natural support is to include compression in the file system \cite{10.1145/2501620.2501623,ntfs,e2compr,gao2019erofs,264858}. 
For example, Gao et al. \cite{gao2019erofs} proposed a compressed read-only file system.
Hu et al. \cite{hu2019qzfs} proposed to exploit hardware compression accelerators (HCA) on the host side. 
Due to increased CPU cost and duplicate compression, HCA-assisted CPU rarely combines with CCSDs.
This method can not be directly reused to control the temperature of CCSDs either.
Compression has also been integrated into databases \cite{sears2008rose}.
Waltz differs from these studies as it focuses on solving novel thermal challenges through hardware-software cooperative compression, which overcomes the limitations of both the host and device sides.

\noindent
\textbf{Device-side compression.}
In recent years, device-side compression has begun to receive much attention \cite{zuck2014compression,zhanghotstorage20Rethink,hu2019qzfs,7843601,10.1145/3126511}.
Zhang et al. \cite{194420} proposed a device-side compression technique to reduce write pressure and extend the lifetime of flash memory.
Several enterprise SSD storage system vendors, e.g., Scaleflux \cite{scaleflux} and Pure Storage \cite{purestorage}, have released storage devices with embedded compression engines.
Song et al. \cite{song2023f2fs} proposed to use the host side to guide the CCSD for selective compression.
However, the in-flight shutdown caused by the CCSD's overheating problem may cause catastrophic damage.
We explore a new way of solving the overheating problem from the perspective of co-compression with the novel temperature status of CCSDs for scheduling compression tasks.
Based on the flexible co-compression design, we propose Waltzs and Waltzp schemes with different trade-off points to serve compressed space or performance needs. 

\noindent
\textbf{Space saved by compression.}
SC and GC are important sources of WAF.
Improving SC and GC efficiency has been widely researched \cite{Park2021Lightweight,Lange2022Offline}.
First, several studies have attempted to migrate the GC overhead from the storage device to the host side \cite{196194}.
The host side manages the flash space directly and performs basic management activities \cite{273945}.
Second, other studies have proposed methods to improve GC efficiency \cite{227784,8782508,7177770}. 
For example, Kim et al. \cite{227784} proposed a mechanism to reduce data movement in GC by distinguishing data with similar lifetimes based on the program context.
Third, some studies propose methods to trigger background SC \cite{190595,5762711,202247,10.1145/1176887.1176911,10.1145/2741948.2741949}.
Different from the above studies, we innovatively explore the usage of the saved space by compression to reduce the cleaning cost of LFSs and show the benefits of compression.


\section{Conclusion}\label{sec:conclusion}
In this paper, we study the characteristics of the host-side and device-side data compression schemes and analyze their benefits and drawbacks. 
The results demonstrate that the host-side compression suffers from high (de)compression overhead and segment cleaning costs, while the device scheme suffers from the CCSD overheating problem.
To achieve efficient data compression with high performance and better temperature control, we propose Waltz, a hardware-software cooperative compression scheme, to control the temperature and improve the performance of CCSD-based storage systems by effectively scheduling (de)compression tasks between CCSD and F2FS at runtime.
Based on Waltz, two schemes are proposed to meet the requirements of different applications.
In addition, we explore Waltz with a case study that exploiting the compressed space for reserved space in F2FS. 
Evaluations on the real CCSD platform show that Waltz mitigates the overheating problem and optimizes the system's performance and WAF. 
In future work, we will explore porting Waltz to other categories of computational SSDs to promote its marketability in the deployment fields.

\bibliographystyle{ACM-Reference-Format}
\bibliography{sample-sigconf}
\end{sloppypar}
\end{document}